\documentclass[12pt]{article}

\usepackage{graphicx}

\title{The spatial string tension and the nonperturbative Debye mass from the Field Correlator Method}
\author{ Yu. A. Simonov  \\
NRC ``Kurchatov Institute'' -- ITEP \\
Moscow, 117218 Russia}

\newcommand{\beq}{\begin{eqnarray}}
 \newcommand{\eeq}{\end{eqnarray}}
\newcommand{\be}{\begin{equation}}
 \newcommand{\ee}{\end{equation}}

\def\fun#1#2{\lower3.6pt\vbox{\baselineskip0pt\lineskip.9pt
\ialign{$\mathsurround=0pt#1\hfil ##\hfil$\crcr#2\crcr\sim\crcr}}}

\newcommand{{\SD}}{\rm SD}

\newcommand{{\Mc}}{\mathcal{M}}

\newcommand{\ver}{\mbox{\boldmath${\rm r}$}}

\newcommand{\vep}{\mbox{\boldmath${\rm p}$}}

\newcommand{\vez}{\mbox{\boldmath${\rm z}$}}

\newcommand{\lan}{\langle}
\newcommand{\ran}{\rangle}

\begin{document}
\maketitle
\begin{abstract}

The phenomenon of the almost linear growth of the square root of spatial string tension $\sqrt{\sigma_s(T)}= c_{\sigma} g^2 T$ and of the Debye mass $m_D(T)$ was found both in lattice and in theory, based on the Field Correlator Method (FCM). In the latter the string tension (both spatial and colorelectric) is expressed as an integral of the two gluon Green's function calculated with the same string tension: $\sigma= \int G^{(2g)}_{\sigma}$. This relation allows to check the selfconsistency of the theory and at high T it allows also to calculate $c_{\sigma}$ and hence $\sigma_s$. We calculate below in the paper the corresponding coefficients $c_{\sigma}$ and $c_D$ numerically in the FCM method and compare the results with lattice data finding a good agreement. This justifies the use of the FCM in the space-like region and in high T thermodynamics without extra parameters.

\end{abstract}

\section{Introduction}

The phenomenon of confinement in QCD was explained in the framework of the Field Correlator Method (FCM) \cite{1,2,3,4,5,5*}, both qualitatively and quanitatively, via the vacuum field correlators  of the colorelectric (CE) and the colormagnetic (CM) fields $E_i^a,H_i^a$, and at the temperature $T=0$ the behavior of all physical quantities is expressed  via the only nonperturbative parameter -- the string tension, $\sigma_E=\sigma_H= \sigma$. The resulting field correlators $D^E(x),D^H(x)$ are the so-called gluelump Green's functions, which define all confining QCD dynamics. These gluelumps have been calculated in good agreement between the FCM \cite{6} and the lattice data \cite{7}, while $D^E(x),D^H(X)$ were also studied in detail at $T>0$ on the lattice \cite{5}. It was found that
the situation is drastically changing at $T>0$, where both $\sigma_E(T)$ and $\sigma_H(T)= \sigma_s(T)$ have a different behavior. Moreover  $\sigma_E(T)$
disappears above $T=T_c$, while in contrast to that $\sigma_s(T)$ grows almost linearly at large $T$, as it was
found on the lattice \cite{4,7*,8,9} and supported by the studies in the framework of the FCM \cite{10,11,12}. This dependence contains the coefficient, denoted as $c_{\sigma}$  \cite{8,9},

\be
\sqrt{\sigma_s(T)}= c_{\sigma} g^2(T) T,
\label{1}
\ee
defined numerically in the lattice calculations \cite{8,9} in the case  $N_c=3, N_f=0$ as
\be
c_{\sigma}= 0.566\pm 0.013.
\label{2}
\ee
For the $SU(2)$ theory a similar lattice calculation \cite{13,13*} has given a smaller value,
\be
c_{\sigma}^2= (0.136 + 0.011),~~ c_{\sigma}= 0.369\pm 0.015.~~
\label{3}
\ee

On the theoretical side the linear temperature growth of $\sigma_s(T)$ was derived in the framework of FCM \cite{10,11,12}, where the basic notion is the colormagnetic (CM) field correlator
$D^H(z)$, which can be expressed via the  gluelump Green's function $G(x.y|z)$ and in this case the (two-gluon) gluelump  is the string triangle, created by two gluons and the adjoint spectator point. The exact theory of these gluelumps was given in \cite{6,10,11} and will be discussed in the next section in the case of the CM correlators. Note that the gluelump Green's
functions define the field correlators in the self-consistent way, without fitting  parameters.  In the case of the CM field  the resulting $\sigma_s(T)$ has exactly the same form as in (\ref{1})
\cite{12} with $c_{\sigma}$, expressed via the gluelump Green's function integral $G(0,0|z)$,
\be
c_{\sigma}^2= \frac{N_c^2 -1}{4} \int d^2 z G(0,0|z).
\label{4}
\ee
However, in the FCM the numerical values of $c_{\sigma}$ were not calculated before and therefore the basic role of the field correlators in the case of the CM confinement was not made certain.

It is the purpose of present paper to calculate the numerical value of $c_{\sigma}$ and compare it with the lattice data. In addition we complete this paper with the comparison of these results with additional information on $\sigma_s$ from the Debye screening length (the Debye mass), which in the FCM is also expressed via the $\sigma_s (T)$ \cite{14,15}. As was shown in \cite{15}, the
nonpertirbative Debye mass is expressed through  $\sigma_s$ as
\be
m_D (T)= 2.06 \sqrt{\sigma_s(T)}= 2.06 c_{\sigma} g^2(T/T_c) T,
\label{5}
\ee
where $g^2(t)$ is defined in the standard way (see Appendix A). In section 3 we shall compare the resulting values of $c_{\sigma}$, following  from the lattice calculation of $m_D$ \cite{16,17},  with those, obtained  in the lattice measurements of the $\sigma_s$ \cite{8,9}, and finally with $c_{\sigma}$, calculated here via the gluelump Green's function, as in (\ref{4}).
As will be seen , the results, obtained within the FCM,   provide the values
of $c_{\sigma}$ in the same ballpark with lattice data within the accuracy of $(10-15\%)$.

\section{The spatial string tension in the FCM}

In Introduction we have mentioned that the spatial Green's function, which defines the $\sigma_s$, is the Green's function of two gluons and a straight gluonic Wilson line (the parallel transporter), all connected by three confining strings, and the string tension is proportional to the integral of this Green's function in the $3d$ space, where one of three space coordinates can be taken as an evolution parameter (the Euclidean ``time"). To calculate $\sigma_s(T)$ one can start with the technic, developed in \cite{10,11,12} for
$D^E(z), D^H(z)$, which allows to express it via the two-gluon  Green's function: $G^{(2g)}_{4d} (z) =  G_{4d}^{(g)}\otimes G_{4d}^{(g)}$, where two gluons
interact nonperturbatively, and later we shall neglect  the spin interactions in the first approximation.

\be (-D^2)^{-1}_{xy}=\left\lan x\left|\int^{\infty}_0 dt e^{tD^2(B)}\right|y\right\ran=
\int^{\infty}_0dt(Dz)^w_{xy}e^{-K}\Phi(x,y), \label{6}
\ee
where
\be
K=\frac{1}{4}\int^s_0d\tau \left(\frac{dz_\mu}{d\tau} \right)^2,
~~\Phi(x,y)=P \exp ig\int^x_yB_{\mu}dz_{\mu},\label{7}
\ee
and a winding path measure is
\be
(Dz)^w_{xy}=\lim_{N\to \infty}\prod^{N}_{m=1}\frac{d^4\zeta(m)}{(4\pi\varepsilon)^2}
\sum^{\infty}_{n=0,\pm1,..}
\int\frac{d^4p}{(2\pi)^4}e^{ip(\sum\zeta(m)-(x-y)-n\beta\delta_{\mu
4})}.
\label{8}
\ee

The starting point  for the gluon propagator $G^{(g)}_{4d}$ is the integration in the 4-th direction in (\ref{6}) with the exponent $K_4 = \frac14 \int^s_0 d\tau \left( \frac{d z_4}{d\tau}\right)^2,$ which gives for the spatial loop with $x_4=y_4$,
\be
J_4 \equiv \int (Dz_4)_{x_4x_4} e^{-K_4} = \sum_{n=0, \pm 1,...} \frac{1}{2 \sqrt{\pi s}} e^{-\frac{(n\beta)^2}{4s}} =
\frac{1}{2\sqrt{\pi s}} \left( 1+ \sum_{n=\pm 1,\pm 2}
e^{-\frac{(n\beta)^2}{4s}}\right).\label{9}
\ee
The  second term in (\ref{9}) at large $T\gg \frac{1}{2\sqrt{s}}$ yields
$2\sqrt{\pi s} T$,    which gives $J_4 =\frac{1}{2\sqrt{\pi s}}+T$.

As  a result the $4d$ gluon propagator reduces to the  $3d$ one,
\be
G_{4d}^{(g)}(z) = T G_{3d}^{(g)}(z) + K_{3d} (z),\label{10}
\ee
where $K_{3d}(z)$ does not depend on $T$. In what follows we consider only the first term in
(\ref{10}), assuming the limit of large $T$.  Substituting this term in  the  general expression for $D^H(z)$, obtained in \cite{12}, one has

\be
D^H(z) = \frac{g^4(N^2_c-1) }{2} \lan G^{(2g)}_{4d} (z)\ran  \to
\frac{g^4(N^2_c-1)T^2}{2} \lan G^{(2g)}_{3d} (z)\ran, \label{11}
\ee
where $G^{(2g)}_{3d}$ is the two-gluon Green's function in $3d$ space with all interactions between gluons taken into account,

\be
\lan  G^{(2g)}_{3d}\ran = \lan G^{(g)}_{3d}(x,y) G^{(g)}_{3d}(x,y)\ran_B. \label{12}
\ee

In terms of the gluelump phenomenology, studied in \cite{6,7}, the expression (\ref{11}) is called the two-gluon gluelump, which was computed on the lattice \cite{7} and analytically in
\cite{6}. In our case we are interested in the $3d$ version of the corresponding Green's function. Choosing in $3d$ the $x_3\equiv t$ axis as the
Euclidean time, we proceed, as in \cite{7}, exploiting the path integral technic \cite{10,11,12}, which yields

\be
G^{(2g)}_{3d} (x-y) = \frac{t}{8\pi} \int^\infty_0 \frac{d\omega_1}{\omega_1^{3/2}} \int^\infty_0
\frac{d\omega_2}{\omega_2^{3/2}} (D^2z_1)_{xy}(D^2z_2)_{xy}
e^{-K_1(\omega_1)-K_2(\omega_2)-Vt}, \label{13}
\ee
where $V$ includes the spatial confining interaction between the three objects: gluon 1, gluon 2, and  the fixed straight line of the parallel transporter, which makes all construction
gauge invariant (see \cite{6,15} for details). In (\ref{A2.3}) $t=|x-y|\equiv |w|;$ and finally

\be
\sigma_s (T) =\frac{g^4(N^2_c-1)T^2}{4} \int \lan G^{(2g)}_{3d} (w)\ran d^2w.\label{14}
\ee

Constructing in the exponent of (\ref{13}) the three-body Hamiltonian in the $2d$ spatial coordinates,
\be
H(\omega_1, \omega_2) = \frac{ \omega_1^2+ \vep^2_1}{2\omega_1}+  \frac{
\omega_2^2+ \vep^2_2}{2\omega_2}+ V(\vez_1, \vez_2), \label{15}
\ee
one can rewrite (\ref{13}) as follows (see \cite{12}),
 \be
 G^{(2g)}_{3d} (t) = \frac{t}{8\pi} \int^\infty_0 \frac{d\omega_1}{\omega_1^{3/2}} \int^\infty_0
\frac{d\omega_2}{\omega_2^{3/2}} \sum^\infty_{n=0} |\psi_n (0,0)|^2 e^{-M_n
(\omega_1,\omega_2) t}.\label{16}
\ee
Here $\Psi_n(0,0)\equiv \Psi_n (\vez_1,\vez_2)|_{\vez_1=\vez_2=0}$, and $M_n$  is the eigenvalue of $H(\omega_1,
\omega_2)$. The latter was studied in \cite{6} in three spatial coordinates. For our purpose here we only mention that $ G^{(2g)}_{3d}  (z)$ has the dimension
of the mass squared and therefore the integral (\ref{14})  is  dimensionless. Hence, one obtains $\sqrt{\sigma_s (T)} = g^2 T  c_{\sigma}$, as it was stated in (\ref{1}), where
 \be
 c^2_\sigma = \frac{(N_c^2-1)}{4} \int d^2 w \lan G_{3d}^{(2g)} (w)\ran. \label{17}
\ee

At this point we realize that the main problem is the calculation of the eigenvalues and the eigenfunctions  of the gluelump Hamiltonian $H(\vep_1,\vep_2)$ in the $3d$ space-time, which can be written as
\be
 H(\vep_1,\vep_2)= \frac12 (\omega_1 + \omega_2) + \frac12 \left(\frac{\vep^2_1}{\omega_1} + \frac{\vep^2_2}{\omega_2}\right)+ \sigma ( |\ver^2_1|+|\ver^2_2|+|\ver_1-\ver_2|^2) + ({\rm spin-dep. terms}),
\label{18}
\ee
and the relativistic eigenvalues $M_n(\omega_1,\omega_2)$ and the eigenfunctions $\psi_n(\ver_1,\ver_2)$ are defined via the gluelump Green's function (\ref{16}) and finally via  $c^2_\sigma$
(\ref{17}).  As it is seen in (\ref{16}), the result can be written as a sum of positive terms,
 \be
 c^2_{\sigma} = \sum^\infty_0 c_{\sigma}^2(n).
 \label{19}
\ee
The first terms with $n=0,2$, which is the lower bound of $c^2_{\sigma}$ is calculated in Appendix A, neglecting the spin-dependent terms. It gives
 \be
 c^2_{\sigma}(n=0)= 0.186 , c^2_{\sigma}(n=2) = 0.024,  c^2_{\sigma} > 0.21.
\label{20}
\ee
Thus one can see that our result, $c_{\sigma}({\rm theor}) > 0.46 $, is in good agreement with the lattice data \cite{8,9} in (\ref{2}): $ c_{\sigma}= 0.566\pm 0.013$. Moreover, it is also in agreement with the $N_c=2$ lattice data \cite{13}, given in (\ref{3}). One should also note that the inclusion of the spin dependent terms for the lowest gluelump tends to increase the value of $c^2_{\sigma}$, as it follows from the analysis of \cite{6}. In this way one can come to the conclusion that the gluelump calculation
of the spatial string tension is in agreement with the corresponding lattice data on $\sigma_s$.

We now turn to the Debye mass problem, which is directly connected with $\sigma_s$ and gives an additional information on this topic.

\section{The nonperturbative Debye mass and the spatial string tension}

As it was argued in \cite{14,15}, the Debye mass $m_D(T)$ is produced by the extension of the quark-antiquark
Wilson loop by exchange of the gluon in the spatial direction. The resulting increase of the Wilson loop surface $\Delta S$  occurs mostly in the $3d$ space  and provides an additional factor
in the Green's function: $f= \exp(-\sigma_s({\rm adj}) \Delta S)$, which enters in the exchanged gluon Green's function and defines its mass $m_D(T)$. Note that the resulting object is the one-gluon gluelump and using the same technique,  as in the previous section, one can predict $m_D(T)$ nonperturbatively. This technique was already used in \cite{14} to calculate the so-called meson and glueball screening masses and what was done in \cite{15} is actually the calculation of the one-gluon gluelump screening mass $m_D(T)$, which was found on the lattice in the form \cite{16,17},
\be
m^{\bf Latt}_D(T)= A_{N_f} \sqrt{1 + \frac{N_f}{6}} g(T) T
\label{21}
\ee
Here  $A_0= 1.51, A_2= 1.42$. From the gluelump approach \cite{15} one obtains $m_D(T)= 2.06 \sqrt{\sigma_s(T)}$ and this result with
$\sqrt{\sigma_s(T)}= c_{\sigma} g^2(T) T$ from (\ref{1}) can be compared with the lattice data for $m_D(T)$ from \cite{17},  given by (\ref{21}): $m^{\bf Latt}_D(T)= 1.98 T$. This value can be compared with the gluelump result (shown above), yielding $m_D(T)= 3.53 c_{\sigma} T$. Thus the lattice data from \cite{17}, together with the lattice data of \cite{8,9},  $c_{\sigma}= 0.56$ are in good agreement with the gluelump prediction from \cite{15}. A more detailed comparison with the lattice data \cite{17}, done in \cite{15}, shows a good agreement  both for $N_f=0$ and $N_f=2$. In this way one can make sure that the nonperturbative spatial dynamics is correctly predicted by the FCM in its gluelump Green's function formalism.
One can notice that the form used in the lattice studies \cite{16,17} as in (\ref{21}) in comparison with our gluelump form
above suggests that the parameter $A_{N_f}$ should be proportional to $g(T)$ and a good agreement of both forms implies
the weak T dependence of the coupling constant $g(T)$.

\section{Conclusions}

in our paper in the framework of the FCM  we have discussed the confinement mechanism for $T>T_c$, which is defined by the colormagnetic field correlators $D^H (Z)$ and the spatial string tension $\sigma_s(T)$. It was shown that the resulting physical picture is rather specific and strongly connected with the transverse motion
of the colored objects, since purely longitudinal motion is associated only with perturbative QCD interaction, which  decreases at large T. Therefore the nonperturbative effects are strongly $T$ --
dependent because  the transverse motion is generated by temperature -- for example,  $m_D$ is growing with $T$ almost linearly.
As a result in the FCM the spatial string tension is defined  by the $3d$ gluelump Green's function $G(w)$  and the corresponding $3d$ gluelump mass is expressed via the $3d$ string tension, so that
finally we obtain in (\ref{16}) the expression  $\sigma_s= c^2_{\sigma} g^4 T^2$ which is defined only by the constant $c_{\sigma}$. Therefore in contrast with the colorelectric string tension $\sigma_E$, which is the basic independent QCD parameter, the $\sigma_s$ (or rather $c_{\sigma}$) can be computed within the theory and it is very important for the theory to test in this way its self-consistency. Here we have provided  this test within  our theory of confinement, based on the FCM, and for  $\sigma_s$ and $c_{\sigma}$ we found
 a good agreement with lattice data, providing in this way good arguments in favor of its self-consistency. An additional
information in favor of the FCM approach was given above, comparing lattice and the FCM predictions  for the Debye mass, as it was done in \cite{15}.
Even more important  role of the spatial string tension may be  in the high $T$ thermodynamics where
in the framework of FCM it provides the basic nonperturbative contribution to the pressure and other observables \cite{18} in good agreement with the lattice data.

The author is grateful to A. M. Badalian for useful discussions and to N. P. Igumnova for help in preparing the manuscript.

\section*{ Appendix A1. Two-loop expression for $g^{-2} (t)$}.

 \setcounter{equation}{0} \def\theequation{A1.\arabic{equation}}

This expression has the standard form as a function of $t= \frac{T}{T_c}$ :
\be
g^{-2}(t)= 2 b_0 ln \left(\frac{t}{L_{\sigma}}\right) + \frac{b_1}{b_0} ln(2 ln\left(\frac{t}{L_{\sigma}}\right),
\label{A1.1}
\ee
where
\be
b_0= \left(\frac{11}{3} N_c -\frac23 N_f\right) \frac{1}{16 \pi^2},~~ b_1= \left(\frac{34}{3} N^2_c -\left(\frac{13}{3} N_c - \frac{1}{N_c}\right) N_f\right) \frac{1}{(16\pi^2)^2}.
\label{A1.2}
\ee

Here $L_\sigma= \frac{\Lambda_\sigma}{T_c}= 0.104 +/- 0.009$ \cite{8,9}.

\section*{Appendix A2. Numerical calculation of $c^2_{\sigma}$ in the lowest approximations}

 \setcounter{equation}{0} \def\theequation{A2.\arabic{equation}}

We calculate here two lowest eigenvalues and eigenfunctions of the Hamiltonian (\ref{18}), which enter
 in the expression for the $c^2_{\sigma}$ in (\ref{16}). This Hamiltonian without spin-dependent terms can be written in the equivalent oscillator form,
 \be
 H= \frac{\omega^2_1 + \vep^2_1}{2\omega_1} + \frac{\omega^2_2 + \vep^2_2}{2\omega_2} + \frac{\sigma^2\ver^2_1}
{2 \nu_1} + \frac{\sigma^2 \ver^2_2}{2 \nu_2} + \frac{\sigma^2 (\ver_1- \ver_2)^2}{2 \nu_3} + \frac{\nu_1 + \nu_2 + \nu_3}{2}.
\label{A2.1}
\ee
Here we have used the property $\sigma |\ver|= \min\left(\frac{\sigma^2 \ver^2 + \nu^2}{2 \nu}\right)$, so that
minimizing the eigenvalues of (\ref{A2.1}) in the variables $\nu_i,~)$, we obtain the eigenvalues and eigenfunctions of the Hamiltonian with a good accuracy $\sim 5\%$.
Then for $\omega_1 = \omega_2 = \omega$ and $\nu_1= \nu_2=\nu$ one obtains the lowest eigenvalue for n=0,

\be
 M_0= \omega + \frac{\sigma}{\sqrt{\omega\nu}}\left(1 + \sqrt{\frac{\nu_3 +2\nu}{\nu_3}}\right) + \frac{2\nu+\nu_3}{2}.
\label{A2.2}
\ee
The conditions of minima $\frac{\partial M_0}{\partial z_i}= 0$ with $z_i= \omega_i,\nu_i$ yields the
final result with notation $\omega_i(0),\nu_i(0)$ for the extremal values.

\be
 \omega_1(0)= \omega_2(0)= 1.29 \sqrt \sigma,~ \nu_0= 0.79 \sqrt \sigma,~ \nu_3(0)= 1.25 \nu_0,~
 \min (M_0)=4.95 \sqrt \sigma.
 \label{A2.3}
 \ee
From the oscillator wave functions it easy to get the factor $|\psi(0,0)|^2= 1.61 \sigma^2$ and
to calculate the integrals over $d\omega_1 d\omega_2$ in \ref{16}, expanding $M_0(\omega_1,\omega_2)$ near the stationary points in (\ref{A2.3}) up to the second order in $\omega_i-\omega_i(0)$ and denoting the second derivative of $M_0$ as $M"(\omega_0)$. Then for the integral in (\ref{17})
one has
\be
\int d^2 w G(w)= \frac{2\pi |\psi(0,0)|^2}{M^2_0 \omega^3_0 M"(\omega_0)}.
\label{A2.4}
\ee
Inserting the stationary values from (\ref{A2.3}) and the second derivative at the stationary point
$M"(\omega_0)= 0.51 \sigma^{-1/2}$, one finally obtains

\be
 \int d^2 w G(w)= \frac{2\pi 0.228 |\psi|^2}{\sigma M^2_0}= 0.093,~
c^2_{\sigma} = \frac{N^2_c-1}{4} \int d^2 w G(w) = 0.186 ~(N_c=3).\label{A2.5}
\ee
In a similar way one can calculate the contribution of the $n=2$ term in the (\ref{19}), which yields approximately
$c^2_{\sigma}(n=2)= 0.019$ and for the sum of two terms with $n=0,2$   $c^2_{\sigma} = 0.205$ for $N_c=3$ which is the lower bound. However as will be seen below the most important corrections appear when one estimates the accuracy of the replacement
of the original linear confinement Hamiltonian (\ref{18}) by the oscillator Hamiltonian (\ref{A2.1}). To get an idea of this effect we can estimate the ratio of the integral in (\ref{A2.5}) which we denote as $I_{\rm osc}$ and the corresponding integral
for the real (linear) interaction $I_{ \rm lin}$. To simplify matter we replace $|\psi(0)|^2$ and $M_0$ of the gluelump system by the  simple two gluon system connected by the linear or oscillator interaction and write approximately
\be
R= \frac{I_{\rm lin}}{I_{\rm osc}} \approx \frac{|\psi_{\rm lin}(0)|^2 M^2_{\rm osc}}{|\psi_{\rm osc}(0)|^2 M^2_{\rm lin}}.
\label{A2.6} \ee
For the two-gluon system with linear confining interaction the spectrum and wave functions are well known \cite{19,20,21}:
\be
M_n= 4 \sqrt{\sigma} (\frac{a(n)}{3})^{3/4},~ |\psi_{\rm lin}(0)|^2= \frac {\sigma M_0}{16 \pi},
\label{A2.7} \ee
where for the ground state $n=0$ $a(0)= 2.338$.
Inserting for the linear and oscillator potentials the resulting values $M_{\rm lin}= 3.31 \sqrt {\sigma}$, $ M_{\rm osc}= 3.59 \sqrt {\sigma}$ and
$ |\psi_{\rm lin}(0)|^2= 0.065 \sigma^{3/2}$, $ |\psi_{\rm osc}(0)|^2= 0.043 \sigma^{3/2}$, one obtains the approximate ratio which estimates  the effect of the replacement by oscillator interaction
\be
R=  \frac{c^2_{\sigma}({\rm lin})}{c^2_{\sigma}({\rm osc})} \approx 1.82;~ c_{\sigma}({\rm lin}) \approx  1.35 c_{\sigma}({\rm osc}).
\label{A2.8} \ee
As a result using the $n=0$  oscillator value of $c^2_{\sigma}$ in (\ref{A2.5}) we obtain  the linear confinement coefficient $c_{\sigma} \approx 0.582 $ which agrees well with the lattice value $ 0.566 \pm 0.013$ from \cite{8,9}.
A more accurate calculation of the $c_{\sigma}({\rm lin})$ is possible with the solution of the linear integral equations for the
gluelump Green's functions as it was done in \cite{6} for the gluelump masses.


\begin{thebibliography}{99}

\bibitem{1}
H. G. Dosch and Yu.A.Simonov, Phys. Lett. {\bf  B 205}, 339 (1988).

\bibitem{2}
Yu. A. Simonov, Phys. Usp. {\bf 166}, 337 (1996), arXiv: hep-ph/9709344; D. S. Kuzmenko, V. I. Shevchenko, and Yu. A. Simonov, Phys. Usp. {\bf 47}, 1 (2004),
arXiv: hep-ph/0310190.

\bibitem{3}
A. Di Giacomo, H. G. Dosch, V. I. Shevchenko, and Yu. A. Simonov, Phys. Rept. {\bf 372}, 319 (2002), arXiv: hep-ph/0007223.
\bibitem{4}
Yu. A. Simonov, Phys. Rev. {\bf D 99}, 056012 (2019), arXiv: 1804.08946.

\bibitem{5}
 M. D'Elia, A. Di Giacomo, and E. Meggiolaro, Phys. Rev. {\bf D 67}, 114504 (2003), arXiv: hep-lat/0205018.

\bibitem{5*}
Yu. A. Simonov, The colormagnetic confinement in QCD, arXiv: 2203.07850 [hep-ph].

\bibitem{6}
Yu. A. Simonov, Nucl. Phys., {\bf B 592}, 350 (2001), arXiv: hep-ph/0003114.

\bibitem{7}
I. Jorycz and C. Michael, Nucl. Phys. {\bf B 302}, 448 (1988); N. Campbell, I. Joricz, and C. Michael,
Phys. Lett. {\bf B 167}, 91 (1986).

\bibitem{7*}
C. Borgs, Nucl. Phys. {\bf B 261}, 451 (1985); E. Manousakis and J. Polonyi, Phys. Rev. Lett. {\bf 58}, 847 (1987).

\bibitem{8}
G. Boyd et al., Nucl. Phys. {\bf B 469}, 419 (1996), arXiv: hep-lat/9602007.

\bibitem{9}
F. Karsch, E. Laermann, and M. Lutgemeier, Phys. Lett. {\bf B 346}, 94 (1995), arXiv: hep-lat/9411020.

\bibitem{10}
Yu. A. Simonov, Phys. Atom. Nucl, {\bf 58}, 339 (1995); N. O. Agasian, JETP Lett. {\bf 57}, 208 (1993), ibid. {\bf 71}, 43 (2000).

\bibitem{11}
Yu. A. Simonov, Phys. Atom. Nucl. {\bf 69}, 528 (2006), arXiv: hep-ph/0501182; Yu. A. Simonov and V. I. Shevchenko, Adv. High Energy Phys. {\bf 2009}, 873051 (2009),
arXiv: 0902.1405 [hep-ph].

\bibitem{12}
Yu. A. Simonov, Phys. Rev. {\bf D 96}, 096002 (2017), arXiv: 1605.07060.

\bibitem{13}
G. S. Bali et al., Phys. Rev.  {\bf 71}, 3059 (1993); hep-lat/9306024.

\bibitem{13*}
M. Teper, Phys. Lett. {\bf B 311}, 223 (1993).

\bibitem{14}
E. L. Gubankova and Yu. A. Simonov, Phys. Lett. {\bf B 360}, 93 (1995), arXiv:hep-ph/9507254.

\bibitem{15}
N. O. Agasian and Yu. A. Simonov, Phys. Lett. {\bf B 639}, 82 (2006), arXiv:hep-ph/0604004.

\bibitem{16}
O. Kaczmarek and F. Zantow, Contribution to: Workshop on Extreme QCD, 108-112, arXiv: hep-lat/0512031.

\bibitem{17}
O. Kaczmarek and F. Zantow, Phys. Rev. {\bf D 71}, 114510 (2005), arXiv: hep-lat/0503017.

\bibitem{18}
N. O. Agasian, M. S. Lukashov, and Yu. A. Simonov, Eur. Phys. J. {\bf A 53}, 138 (2017), arXiv: 1701.07959.

\bibitem{19}
Dan La Course and M. G. Olsson, Phys. Rev. {\bf D 39}, 2751 (1989).

\bibitem{20}
W. Lucha, F. F. Schoeberl and D. Gromes, Phys. Rep. {\bf 200}, 127 (1991).

\bibitem{21}
Yu. A. Simonov, Phys. Lett. {\bf B 226}, 151 (1989); Yu. A. Simonov, QCD and Theory of Hadrons, in: ``QCD: Perturbative or
Nonperturbative." Interscience, Singapore, 2000; arXiv: hep-ph/9911237.



\end{thebibliography}
\end{document}